\documentclass[%
 reprint,
 superscriptaddress,
 amsmath,amssymb,
 aps, prb,letters,
]{revtex4-2}
\usepackage{graphicx}
\usepackage{dcolumn}
\usepackage{bm}
\usepackage[mathlines]{lineno}
\usepackage{xcolor}
\usepackage[colorlinks=true, linkcolor=blue, citecolor=blue, urlcolor=blue]{hyperref}
\usepackage{float}
\usepackage{times}

\begin{document}
\preprint{PRB/123-QED}
\pagestyle{empty}
\onecolumngrid
\begin{center}
\vspace*{\fill}
\begin{minipage}{0.85\textwidth} 
    \begin{center}
        \textbf{Notice}
    \end{center}
    This manuscript has been authored by UT-Battelle, LLC, under contract DE-AC05-00OR22725 with the US Department of Energy (DOE). The US government retains and the publisher, by accepting the article for publication, acknowledges that the US government retains a nonexclusive, paid-up, irrevocable, worldwide license to publish or reproduce the published form of this manuscript, or allow others to do so, for US government purposes. DOE will provide public access to these results of federally sponsored research in accordance with the DOE Public Access Plan (\href{https://www.energy.gov/doe-public-access-plan}{https://www.energy.gov/doe-public-access-plan}).
\end{minipage}
\vspace*{\fill}
\end{center}
\newpage
\twocolumngrid

\title{Constraints on magnetism and correlations in RuO$_2$ from lattice dynamics and M\"ossbauer spectroscopy}

\author{George Yumnam}
\email{yumnamg@ornl.gov}
\author{Parul R.~Raghuvanshi}
\author{John D. Budai}
\affiliation{Materials Science and Technology Division, Oak Ridge National Laboratory, Oak Ridge, TN 37831 USA}
\author{Dipanshu Bansal}
\altaffiliation{Current Address: Department of Mechanical Engineering, Indian Institute of Technology Bombay, Mumbai, 400076, India}
\affiliation{Materials Science and Technology Division, Oak Ridge National Laboratory, Oak Ridge, TN 37831 USA}
\author{Lars Bocklage}
\affiliation{Deutsches Elektronen-Synchrotron DESY, Notkestr. 85, 22607 Hamburg, Germany}
\affiliation{The Hamburg Centre for Ultrafast Imaging CUI, 22761 Hamburg, Germany}
\author{Douglas Abernathy}
\author{Yongqiang Cheng}
\affiliation{Neutron Scattering Division, Oak Ridge National Laboratory, Oak Ridge, TN 37831}
\author{Ayman Said}
\affiliation{Advanced Photon Source, Argonne National Laboratory, Lemont, IL USA}
\author{Igor I. Mazin}
\affiliation{Department of Physics and Astronomy, George Mason University, Fairfax, Virginia 22030 USA}
\affiliation{Quantum Science and Engineering Center, George Mason University, Fairfax, Virginia 22030 USA}
\author{Haidong Zhou}
\affiliation{Department of Physics and Astronomy, University of Tennessee, Knoxville, Tennessee 37996, USA}
\author{Benjamin A. Frandsen}
\affiliation{Department of Physics and Astronomy, Brigham Young University, Provo, Utah 84602, USA}
\author{David S. Parker}
\author{Lucas R. Lindsay}
\author{Valentino R. Cooper}
\author{Michael E. Manley}
\author{Rapha\"el P. Hermann}
\email{hermannrp@ornl.gov}
\affiliation{Materials Science and Technology Division, Oak Ridge National Laboratory, Oak Ridge, TN 37831 USA}

\date{\today}

\begin{abstract}
We provide experimental evidence for the absence of a magnetic moment in bulk RuO$_2$, a candidate altermagnetic material,  by using a combination of M\"{o}ssbauer spectroscopy, nuclear forward scattering, inelastic X-ray  and neutron scattering, and density functional theory calculations. Using complementary M\"ossbauer and nuclear forward scattering we determine the $^{99}$Ru magnetic hyperfine splitting to be negligible. Inelastic X-ray and neutron scattering derived lattice dynamics of RuO$_2$ are compared to density functional theory calculations of varying flavors. Comparisons among theory with experiments indicate that electronic correlations, rather than magnetic order, are key in describing the lattice dynamics.\vspace{-0.5em}

\end{abstract}


\maketitle
\setcounter{page}{1}
\pagestyle{plain}

\section*{Introduction}\vspace{-1.125em}

The recent observations of broken Kramer's degeneracy and spin-resolved anomalous Hall effect in antiferromagnets (AFM) has challenged the conventional  Néel theory ~\cite{libor2022altermagnet,chen2014AHEinNCM,igor2021prediction,betancourt2023MnTe}. These, so-called, \textit{altermagnets} simultaneously exhibit broken time-reversal $\left(\mathcal{T}\right)$ and inversion $\left(\mathcal{P}\right)$ ; resulting in non-relativistic spin-splitting in their band structures~\cite{krempasky2024altermagnetic,osumi2024observation, libor2022symmetry,libor2022altermagnet}. This $k$-space dependent spin splitting with zero net-magnetization is of interest for spintronics due to its potential for terahertz spin-current generation and detection with negligible stray fields~\cite{kamil2018thz,tomas2016afms,baltz2018afm,Dal2024AFM}.

\begin{figure*}[t!]
    \centering
    \includegraphics[width=\linewidth]{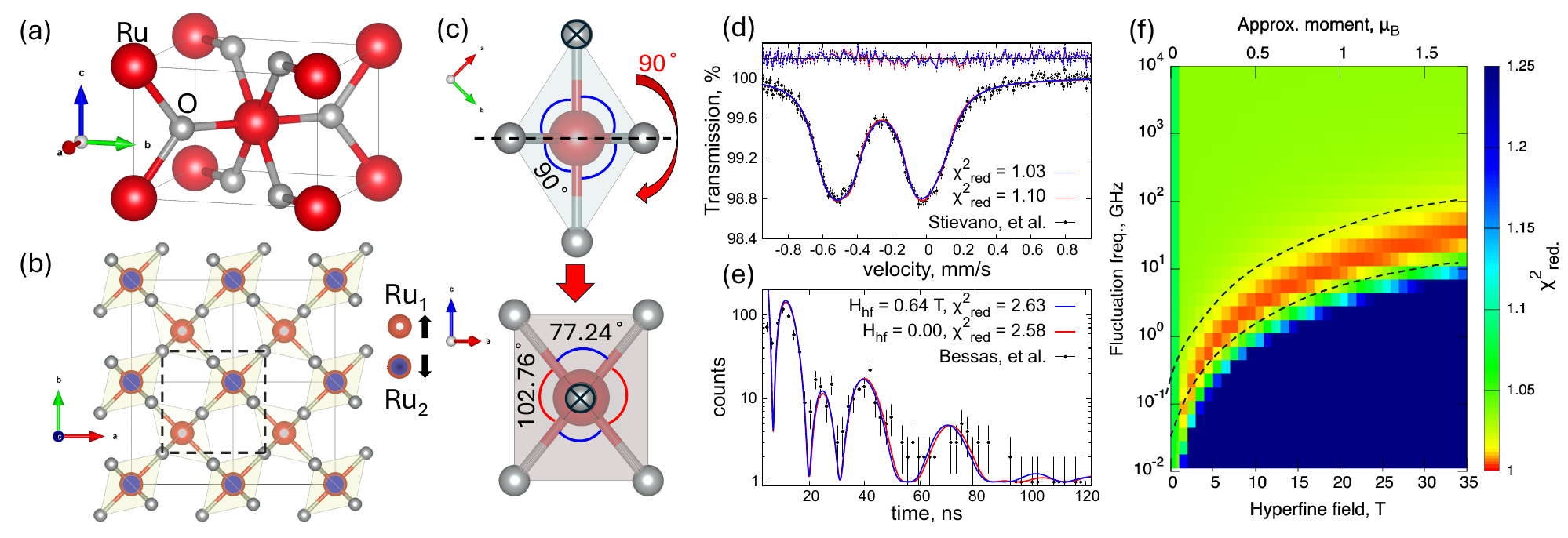}
    \caption{(a) The rutile crystal structure of RuO$_2$ with Ru-atoms in red, and O-atoms in grey. (b) The c-axis projection illustrates the $\left[\mathcal{C}_2 || \mathcal{C}_{4z} t \right]$ symmetry, where Ru$_1$ (spin-up) and Ru$_2$ (spin-down) are associated with RuO$_6$ octahedra oriented differently. The dotted box highlights a unit-cell with a shifted origin that clearly distinguishes Ru$_1$ and Ru$_2$ environments under $\left[\mathcal{C}_{4z} t\right]$ symmetry operation. (c) Representation of the strong distortion in the RuO$_6$ octahedra from two viewing angles. The octahedra consists of 90$^{\circ}$ O-Ru-O bonds along with 77.24$^{\circ}$ and 102.76$^{\circ}$ angle bonds in the plane of the octahedra. The O-atom marked with $\otimes$ shows the 90$^{\circ}$ transformation between the two viewing angles. (d) $^{99}$Ru M\"ossbauer spectra of RuO$_2$ (data from Ref.~\cite{stievano1999mossbauer}) and (e) nuclear forward scattering data of RuO$_2$ (data from Ref.~\cite{bessas2014nfs}) including the fits with-- (blue) and without-- (red) hyperfine field ($H_{\mathrm{hf}}$). (f) Map of reduced $\chi^2$ as a function of the spin relaxation frequency and hyperfine field ($H_{\mathrm{hf}}$). The red-region ($\chi^2_{\mathrm{red.}} \simeq 1$) represents best fit (dashed lines are a guide to the eye).}
    \label{fig1}
\end{figure*}

Ruthenium dioxide (RuO$_2$) is one of the most extensively explored materials with respect to potential altermagnetic characteristics ~\cite{libor2020che,Feng2022anomalous,libor2023chiral,zhou2024crystal,olena2024obsTR,yaqin2024direct}. However, the magnetic structure and size of the magnetic moment on Ru$^{4+}$ ions in bulk RuO$_2$ are debated~\cite{smolyanyuk2024fragility,berlijn2017neutron,kessler2024absence,Hiraishi2024nonmagnetic,jianyu2024absencespinsplit,lovesey2023magstr}, notably due to various reports on thin film properties~\cite{jianyu2024absencespinsplit,Feng2022anomalous,olena2024obsTR}. RuO$_2$ crystallizes in the prototypical rutile structure with the space-group P4$_2$/mnm (136), where each Ru-atom is octahedrally coordinated by six oxygen atoms  that form a distorted RuO$_6$ octahedron~\cite{bolzan1997structural,jacob2000refinement}, see Fig.~\ref{fig1}~(a). The $\mathcal{PT}$ symmetry breaking in RuO$_2$ arises from the two  different orientations of the RuO$_6$ octahedra, see Fig.~\ref{fig1}~(b). In the presence of a spin on Ru, this unique configuration of alternating opposite spins on Ru-atoms, with different orientations of the RuO$_6$ octahedra, is central to the unconventional magnetic and transport properties in RuO$_2$. Altermagnetism requires the magnetic structure of RuO$_2$ to have an anti-parallel coupling of Ru$_1$ and Ru$_2$ magnetic moments polarized along the $\hat{c}-$axis with a propagation wave-vector $\mathbf{q} = (0, 0, 0)$~\cite{libor2022symmetry}. Based on the recently proposed spin-space group~\cite{libor2022symmetry,libor2022altermagnet}, RuO$_2$ would have $\left[\mathcal{C}_2 || \mathcal{C}_{4z}t \right]$, which corresponds to a spin-space two-fold rotation symmetry $\left(\mathcal{C}_2 \right)$ and a real-space four-fold rotation symmetry along the $c$-axis combined with a translation Ru$_1 \xrightarrow{\mathit{t}}$ Ru$_2$ $\left(\mathcal{C}_{4z}t\right)$ as shown in Fig.~\ref{fig1}~(b).

Early single-crystal neutron diffraction measurements observed a small structurally forbidden (100) reflection, which indicates a magnetic reflection, hence suggesting the presence of a magnetic moment in RuO$_2$~\cite{berlijn2017neutron}. This result could only be reproduced with density functional theory (DFT) calculations employing a Hubbard $U$ correction (DFT+$U$) that induced an antiferromagnetic groundstate~\cite{berlijn2017neutron}. A magnetic-moment of $\sim0.05~\mu_{\mathrm{B}}$ was estimated at room-temperature with a magnetic structure corresponding to the P4$_2$/mnm symmetry with oppositely coupled spins on Ru$_1$ and Ru$_2$ aligned along the $c$-axis. A modified model of the magnetic structure in RuO$_2$ was also proposed based on a monoclinic (P$2_1$/c) unit cell, which permits a centrosymmetric magnetic structure that preserves translational symmetry and retains the propagation vector $\mathbf{q} = (0, 0, 0)$~\cite{lovesey2023magstr}. At the same time, non-polarized neutron diffraction could only be fit with a magnetic moment of $\sim 0.23\ \mu_B$, and the lowest magnetic moment observed in the DFT calculations with the modified model was $\sim0.7\ \mu_B$ (or zero, without a $U$-correction). This quantitative discrepancy regarding the size of the presumed magnetic moment was not resolved. 

Subsequent reassessments of bulk RuO$_2$ using single-crystal neutron diffraction revealed that the previously observed intensity of the structurally forbidden (100) reflection~\cite{berlijn2017neutron} is likely a double-scattering artifact~\cite{kessler2024absence,Kiefer2025}. Furthermore, two independent experiments using muon spin rotation ($\mu$SR) spectroscopy found that the magnetic order in RuO$_2$ is either absent entirely or the magnetic moments must be below 10$^{-3} \mu_{\mathrm{B}}$  ~\cite{kessler2024absence,Kiefer2025,Hiraishi2024nonmagnetic}. In contrast, structural analysis revealed that the reduction in dimensionality from bulk to monolayer nanosheets of RuO$_2$ leads to a transformation from a $t-MX_2$ type (tetragonal-like) to a distorted $h-MX_2$ (hexagonal-like) structure~\cite{ko2018understanding}. This transformation is accompanied by an increased strain effect, which could be the origin of several unconventional magnetic and transport properties in 2-dimensional RuO$_2$, which was assumed intrinsic in earlier work~\cite{Feng2022anomalous,olena2024obsTR,ko2018understanding,gregory2022strain,payal2024strain}. A more recent examination indicates that RuO$_2$ exhibits a robust and anisotropic spin Hall effect even in the absence of altermagnetic spin splitting contributions~\cite{wang2025robustanisotropicspinhall}. Similarly, angle-resolved photoemmision spectroscopy (ARPES) studies have demonstrated that depending on sample growth conditions, variations in the spin-splitting can arise via strain or structural distortions. ~\cite{jianyu2024absencespinsplit,olena2024obsTR}. 

Here, we present experimental evidence for the absence of a magnetic moment in RuO$_2$ via a combination of inelastic neutron and X-ray scattering, M\"ossbauer spectroscopy, and nuclear forward scattering techniques. We also employed first-principles calculations of the lattice dynamics based on density functional theory. We used bulk RuO$_2$ single crystal and powders in our experiments to rule out the influence of strain and other distortions arising from low-dimensionality and substrate effects.\vspace{-1.5em}

\begin{figure*}[t!]
    \centering
    \includegraphics[width=\linewidth]{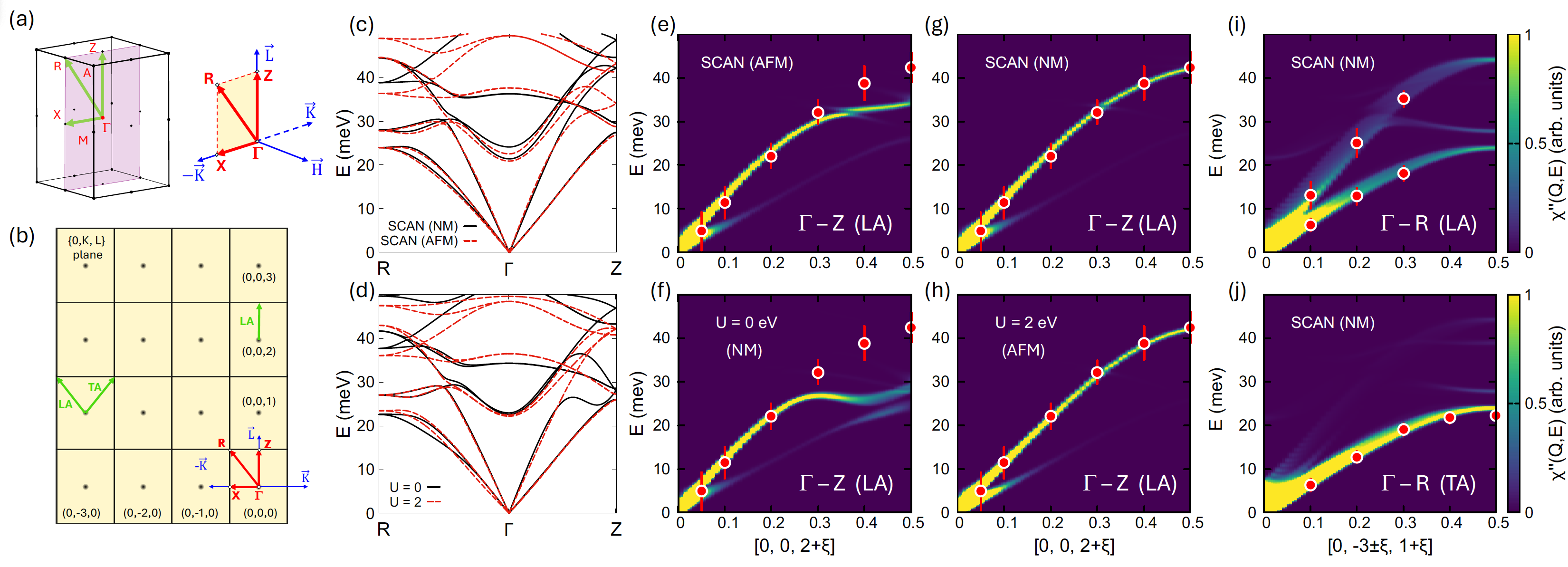}
    \caption{Dynamic structure factor, $S(Q, E)$, of RuO$_2$. (a) The Brillouin zone with high symmetry \textbf{k}-points, drawn next to the $(HKL)$ basis vectors of our IXS experiment. (b) The reciprocal-space map of the $(0KL)$ plane. Green arrows indicate the measured IXS path. Blue and red arrows are the projection of the basis vectors and high-symmetry \textbf{k}-path in the plane. (c) SCAN and (d) DFT+$U$ calculated phonon dispersions. Note the differences in the dispersion of acoustic phonons along $\Gamma-$Z. (e-h) Overlay of the IXS phonons (red circles) on the IXS dynamic structure factor, $S(Q, E)$, computed from SCAN and DFT+$U$ force constants along the $\Gamma-$Z direction for the LA mode reveals that SCAN (NM) is the best model. (i) LA-mode and (j) TA-mode phonons of SCAN (NM) calculation along $\Gamma-$R direction (0, $-$3$\pm\xi$, 1+$\xi$). The error-bars shown in the plot are the FWHM of a Gaussian fit to the IXS constant-Q measurement. Errors of the data points are within the size of the circle. The unit of intensity as shown in the color bar is arbitrary (arb. units).}
    \label{fig2}
\end{figure*}

\section*{Analysis}\vspace{-1.25em}
\emph{M\"ossbauer Spectroscopy and Nuclear Forward  Scattering }
We utilized $^{99}$Ru M\"ossbauer spectroscopy~\cite{kistner1966} to directly probe the magnetic hyperfine field ($H_{\mathrm{hf}}$) at the nucleus,~\cite{long2013mossbauer} This enables the detection of local magnetism, even in the absence of long-range order, both for itinerant and localized moments, and is sensitive to potential transferred hyperfine fields from neighboring atoms. M\"ossbauer spectroscopy also probes the atomic charge state, through the isomer shift ($\delta_{\mathrm{IS}}$), and distortion of the local electronic environment through the quadrupole splitting ($\Delta E_{\mathrm{Q}}$), which probes the electric field gradient (EFG). $^{99}$Ru M\"ossbauer spectra of RuO$_2$ at $\sim 5$~K were previously reported~\cite{stievano1999mossbauer,kistner1966} and were analyzed with a single component fit with $\delta_{\mathrm{IS}} = -0.26(1)$ mm/s and $|\Delta E_{\mathrm{Q}}| = 0.50(1)$ mm/s, indicative of Ru(IV) in a distorted octahedral environment ($\Delta E_{\mathrm{Q}} = 0$ for cubic local symmetry).  The large distortion in the RuO$_6$ octahedra is associated with the asymmetric bond angles between adjacent O-Ru-O bonds in the plane, where neighboring angles are 77.24$^\circ$ and 102.76$^\circ$, respectively, as shown in Fig.~\ref{fig1}~(c). This distortion yields a fairly large EFG. The earlier fits of the M\"ossbauer data did not consider the sign of $V_{\mathrm{zz}}$ and the asymmetry parameter of the EFG, $\eta = \left( |V_{\mathrm{yy}}| - |V_{\mathrm{xx}}| \right)/ |V_{\mathrm{zz}}|$, where $V_{\mathrm{\cdot\cdot}}$ represents principal components of the EFG tensor. The constraint $|V_{\mathrm{zz}}| > |V_{\mathrm{yy}}| > |V_{\mathrm{xx}}|$ mandates that $0 \leq \eta \leq 1$. For $^{99}$Ru, the sign of $V_{\mathrm{zz}}$ and the magnitude of $\eta$ can be determined, even in the absence of magnetic hyperfine splitting, because the nuclear transition occurs between the $I_g=5/2$ ground state and $I_e=3/2$ excited nuclear state. The necessity of using a fairly large $\eta$ is established by $^{99}$Ru nuclear magnetic resonance (NMR) measurements that found $\eta \sim 0.74$~\cite{mukuda1999spinfluct}. Our calculation of the EFG with the r$^2$SCAN meta-GGA functional (SCAN), see methods, in the non-magnetic state yields $V_{\mathrm{zz}}=-10.9\cdot10^{-21}$ V/m$^2$, which corresponds to $\Delta E_{\mathrm{Q}}=0.42$ mm/s, and $\eta=0.426$; $V_{\mathrm{zz}}$ is along the $c$-axis in this calculation. EFG values for other functionals are provided in Table~S1 in \hyperref[suppinfo]{Supplemental Information (SI)}; the only notable discrepancy between models is that for DFT+$U$ = 2 eV in the AFM state, the direction for $V_{\mathrm{zz}}$ is along (110), $i.e.$ between the $a$ and $b$ axes. Note that there is a relative rotation of 90 degrees in the $ab$ plane for the Ru$_1$ and Ru$_2$ EFG, as prescribed by the structure, see Fig.~\ref{fig1}~(b).


We reanalyze the M\"ossbauer spectra of RuO$_2$ from Ref.~\cite{stievano1999mossbauer}, see Fig.~\ref{fig1}~(d),  by considering  $\eta$ in compliance with the $^{99}$Ru-NMR experiment to assess the presence of a possible hyperfine field ($H_{\mathrm{hf}}$), while also considering the sign of the EFG. The first conclusion is that indeed a negative $V_{zz}=-0.54$~mm/s and $\eta\sim0.74$, clearly improve the fit to the data, see comparison of fits with different signs in Fig. S1 of \hyperref[suppinfo]{SI}.  We also used nuclear forward scattering (NFS) of synchrotron radation, which corresponds to time-domain M\"ossbauer spectral data,  from Ref.~\cite{bessas2014nfs}, measured on RuO$_2$ at 10 K. NFS has a better resolution for hyperfine interactions than M\"ossbauer spectroscopy but is insensitive to the sign of the EFG. We  thus  performed a combined fit of the M\"ossbauer spectra and NFS data with the Nexus platform~\cite{nexus2024}, see Fig.~\ref{fig1}~(d, e). We used two models: \emph{(i) Model A: without} $H_{\mathrm{hf}}$ (red) and \emph{(ii) Model B: with} $H_{\mathrm{hf}}$ (blue). Both models employ an asymmetry parameter of $\eta = 0.74$. The model with $H_{\mathrm{hf}}$ is indistinguishable from the model without $H_{\mathrm{hf}}$ within error-bars, which implies that a hyperfine-field is not needed. \emph{Model A} finds a $H_{\mathrm{hf}} = 0.64(15)$~T.  There is no strict proportionality between the hyperfine field and the magnetic moment, but an order of magnitude can be estimated by considering M\"ossbauer data on SrRuO$_3$~\cite{DeMarco2000temperature}. Ru is octahedrally coordinated by O in SrRuO$_3$, and exhibits a hyperfine field of $\sim 33$~T for a 1.6~$\mu_\mathrm{B}$ moment. Accordingly, the best-fit value of 0.64~T in the present case would correspond to a moment of $\lesssim 0.03~\mu_{\mathrm{B}}$ for Ru in $\mathrm{RuO_2}$. 

Given the low moment magnitude, a further option should be considered, namely, whether fast fluctuations of a disordered local moment could explain the observed spectrum. To assess this, we employed the Random Phase Approximation (RPA) formalism on the stochastic theory of relaxation~\cite{Dattagupta1974stochastic} to fit the data for varying spin relaxation frequencies, as used in Ref.~\cite{hermann2006}, on our M\"ossbauer spectrum. The resulting goodness-of-fit is represented by the reduced $\chi^2$ as a function of the fluctuation frequency and hyperfine field, as shown in Fig.~\ref{fig1}~(f). The RPA stochastic-relaxation analysis reveals that the best fits define a region in hyperfine-field $vs$ fluctuation-rate. This region indicates that the larger the potential hyperfine field on Ru, the larger the fluctuation frequency would have to be to exceed GHz frequencies for a 10~T hyperfine field or $\sim$ 0.5~$\mu_{\mathrm{B}}$ moment. Summarizing, considering only M\"ossbauer spectroscopy data, a large moment of $\sim1~\mu_{\mathrm{B}}$ would imply GHz-range fluctuation rates, whereas a static moment would be at most $0.03(1)~\mu_{\mathrm{B}}$, with no improvement in the fits  compared to no moment.

\begin{figure*}[t]
    \centering
    \includegraphics[width=\linewidth]{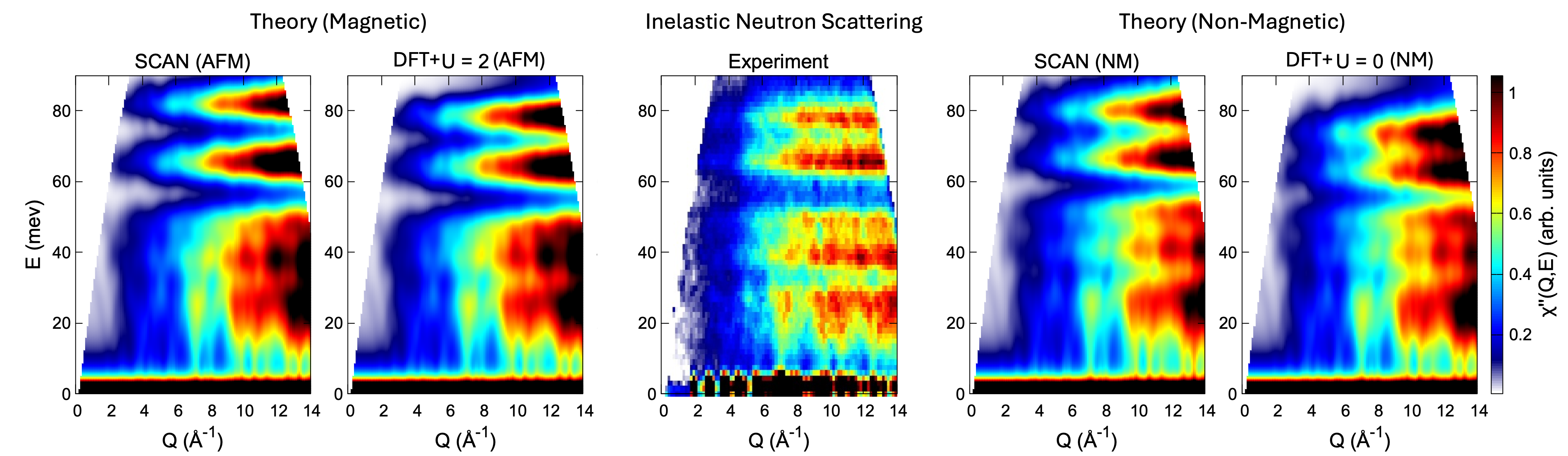}
    \caption{Dynamic susceptibility, $\chi''(Q, E)$, for RuO$_2$ powder spectra measured at ARCS (Experiment, in the middle) and the corresponding OCLIMAX calculated $\chi''(Q, E)$ based on DFT force-constants for SCAN (AFM), DFT+$U$=2 (AFM), SCAN (NM), and DFT+$U$=0 (NM). The intensity represents $\chi''(Q, E)$ in arbitrary units. Calculated intensity is shown with the same scaling for all methods.}
    \label{fig3}
\end{figure*}

\emph{Dynamical structure factor.} Probing lattice dynamics and phonon dispersions of magnetic materials can give meaningful insights into the presence of spins and magnetic order, as spin-lattice interactions can alter force constants. This effect is observable even in weakly magnetic systems, where small magnetic moments and magnetic order arise from spin relaxation or short-range magnetic correlations~\cite{hahn2009influence,YILDIRIM2009425}.

We performed Inelastic X-ray Scattering (IXS) and Inelastic Neutron Scattering (INS) experiments to measure the phonon spectra of bulk RuO$_2$. Subsequently, to assess the sensitivity of phonons to different magnetic configurations, density functional theory (DFT) calculations were used to compute RuO$_2$ phonons in both magnetic and non-magnetic ground states. The transverse (TA) and longitudinal (LA) acoustic phonons along the $\Gamma-$R and $\Gamma-$Z directions were obtained from IXS on single crystal RuO$_2$ around the (0,0,2) and (0,-3,1) Bragg peaks, as illustrated in Fig.~\ref{fig2}~(b). In parallel, DFT-computed phonon band structures were obtained using r$^2$SCAN with the rVV10 nonlocal  correlation functional \cite{r2SCAN} (hereafter SCAN) and PBE-GGA+$U$ (hereafter DFT+$U$) functionals, respectively (Fig.~\ref{fig2}~(c) and (d)). Because DFT+$U$ was shown to be extremely sensitive to the value of $U$~\cite{smolyanyuk2024fragility,parul2025oso2}, we expanded calculations to SCAN, which has been shown to yield reliable predictions of both energetic and structural properties across a range of bonding environments~\cite{r2SCAN}. Calculation details and ground state energies are provided in \hyperref[suppinfo]{SI}. Using the force constants and dynamical matrices computed via DFT, we obtain the dynamic structure factor, $S(Q,E)$, and the Bose-factor corrected dynamic susceptibility, $\chi''(Q,E)$, via the OCLIMAX code~\cite{oclimax2019jctc}, which enables a direct comparison with experimentally measured phonons. Using SCAN, either a non-magnetic (NM) or an antiferromagnetic (AFM) state with a magnetic moment of $\sim$ 0.963 $\mu$B per Ru atom, was obtained depending on whether an initial magnetic moment ($\mu^i$) was specified. Corresponding phonon dispersions are shown in Fig.~\ref{fig2}~(c). Using the DFT+$U$ framework the system exhibits a NM ground state with $U$ = 0 eV, whereas $U$ = 2 eV stabilizes an AFM state with a magnetic moment of $\sim$1.179 $\mu$B on Ru atoms. Corresponding phonon dispersions are shown in Fig.~\ref{fig2}~(d).  There is a striking dependence on the magnetic state and chosen functional for the dispersion of the LA mode in the $\Gamma$-Z direction, as noted earlier in Ref.~\cite{parul2025oso2}. The magnetic moments and lattice parameters predicted using various DFT functionals are presented in Table S2 of the \hyperref[suppinfo]{SI}.


A direct comparison of the IXS phonons (red circles with error-bars) with DFT-based dynamic susceptibilities, $\chi''(Q, E)$, which maps the populated phonons with corrections for the directional polarization is shown in Fig.~\ref{fig2}~(e-h). SCAN (AFM) and DFT+$U$ = 0 (NM) phonons predict inaccurate dispersions, see Fig.~\ref{fig2}~(e) and (f); whereas SCAN (NM) and DFT+$U$ = 2 (AFM) both show very good agreement, see Fig.~\ref{fig2}~(g) and (h). The $\Gamma-$R phonons are properly described by all calculations, as depicted in Fig.~\ref{fig2}~(i) and (j) along with the SCAN (NM) dynamic susceptibility. Summarizing, IXS data for the $\Gamma$-Z LA phonon indicates that non-magnetic SCAN and the AFM state from DFT+$U$=2 both reproduce the scattering intensities.

Inelastic neutron scattering on RuO$_2$ powders yielded the dynamical susceptibility shown in Fig.~\ref{fig3}, center. The measured $\chi''$ can be classified into \textit{(i)} low energy phonons (0 -- 50 meV), which display two features: low lying conical dispersion attributed to the acoustic phonons and two flat optical modes with interconnected branches, and \textit{(ii)} high energy phonons (50 -- 90 meV), which display flat optical bands with few interconnects separated by a gap from the low energy phonons. The calculated $\chi''$  from our four models are shown for comparison; the SCAN (NM) calculated phonons provide an overall best visual match with the experiment. 

\begin{figure}[!b]
    \centering
    \includegraphics[width=0.9\columnwidth]{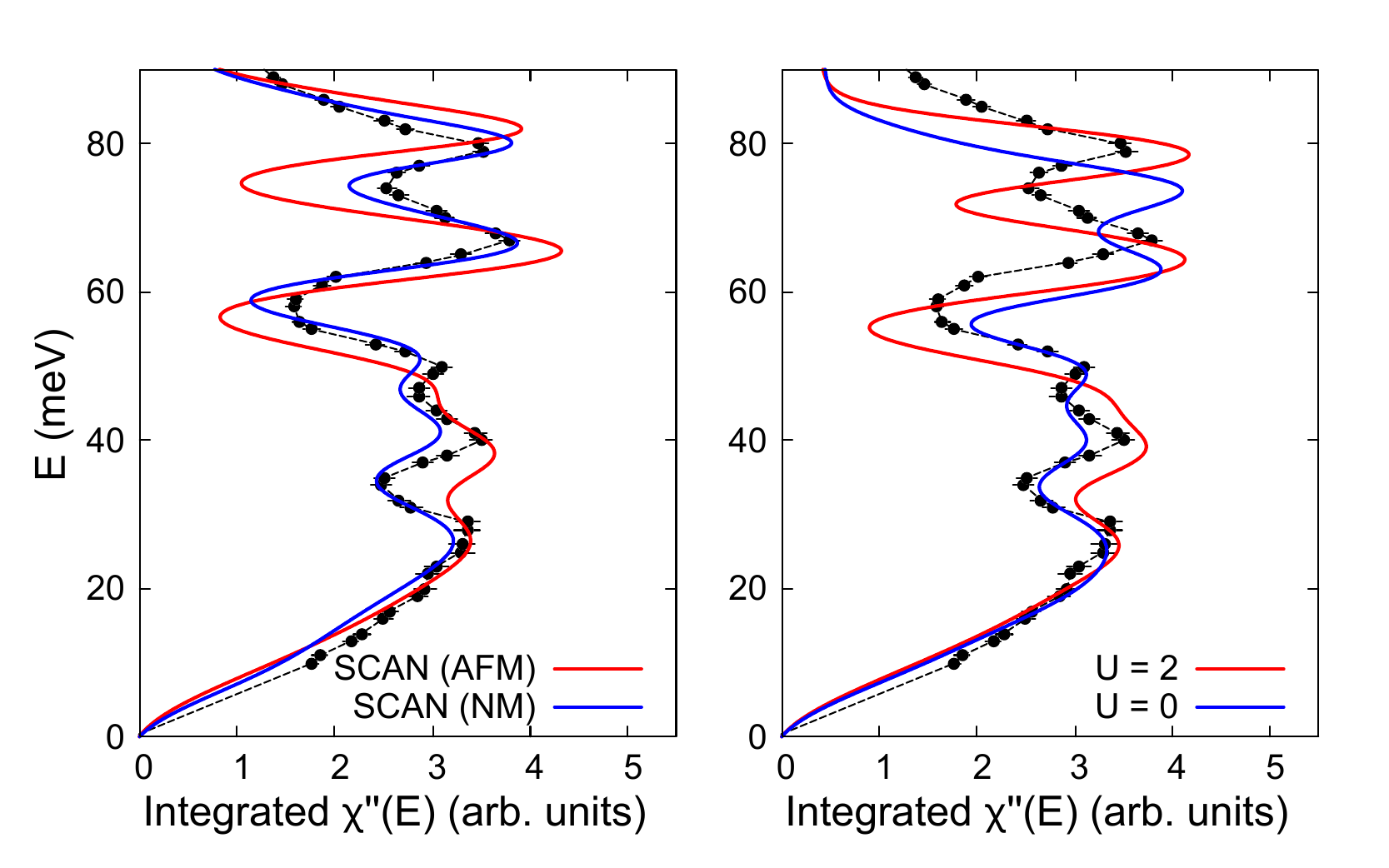}
    \caption{Integrated $\chi''(E)$ for RuO$_2$ from the ARCS measurement is represented by black circles. Correspondingly, the simulated Integrated $\chi''(E)$ for SCAN (AFM), SCAN (NM), DFT+$U$=0 (NM), and DFT+$U$=2 (AFM) is plotted. The simulated data is scaled by the same constant to compare with the experiment.}
    \label{fig4}
\end{figure}

However, a more direct comparison between the experimentally measured phonons and DFT calculations can done by comparing the integrated $\chi''$, i.e., $\int \chi''(Q, E)~dQ$ from $Q = 7 - 12~\mathrm{\AA}^{-1}$ as a function of energy transfer ($E$) as shown in Fig.~\ref{fig4}. 
For this comparison, exact matches in energy should be disregarded as the INS data was obtained at 295~K and are compared to ground-state DFT calculations, i.e., $T$ = 0~K. Considering the overall shape of the models, there is better agreement with SCAN (NM), whereas both DFT+$U$ magnetic orderings fail at higher energies and SCAN (AFM) fails at all energies.\vspace{-1.95em}

\section*{Discussion}\vspace{-1.25em}
Here, we will discuss these observations in light of other recent research on RuO$_2$ magnetism, notably $\mu$SR and neutron diffraction in Ref. \cite{kessler2024absence} and $\mu$SR in Ref. \cite{Hiraishi2024nonmagnetic}. These studies provide strong evidence against, and upper boundaries for, a possible magnetic moment on Ru. A combination of $^{99}$Ru M\"ossbauer spectroscopy and NFS analyzed here likewise indicates that any static Ru moment would be very small, at most 0.03 $\mu_\mathrm{B}$, whereas GHz-frequency fluctuations would be required for any larger moment. Overall, the data can be explained without any magnetic hyperfine splitting on the $^{99}$Ru nuclei. Concerning the possibility of any fluctuating moments, the $\mu$SR data, specifically with applied longitudinal field, also rules out any large uncorrelated moments - which would result in robust muon depolarization even in the presence of a weak applied longitudinal field, in contrast to the observed behavior reported in Ref.~\cite{Hiraishi2024nonmagnetic}. 

A second observation concerns the LA phonons in the $\Gamma$-Z direction, which in our DFT calculations, are highly sensitive to the combination of magnetism and exchange-correlation approximations used. We initially found that the lattice dynamics calculations in the AFM state for DFT+$U$ = 2 provided a good agreement and that the DFT+$U$ = 0 non-magnetic state fails, see Fig.~\ref{fig2}~(f and h), which provides support for a correlated AFM state~\cite{parul2025oso2}. However, non-magnetic calculations with the meta-GGA SCAN functional, see Fig.~\ref{fig2}~(g), reproduce the data equally well, and even better if considering the INS data. This indicates that the presence of strong correlations is critically important for determining the correct behavior of this phonon branch, while magnetism is not necessary. 

The DFT+$U$ method relies on the Hubbard $U$, on-site Coulomb repulsion, that tends to localize Ru 4$d$ electrons. Hence one can arrive at a transition from itinerant to localized electrons with increasing $U$~\cite{smolyanyuk2024fragility}. Static DFT+$U$ treats correlations by enhancing localization that tends to freeze local moments that may exist only transiently in real materials. This method was initially introduced in order to describe magnetism in strongly correlated systems, such as La$_2$CuO$_4$~\cite{anisimov1991band}. It proved to be indispensable for strongly correlated insulators, mostly based on 3$d$-- or $f$--metals, but in some cases also 4$d$ or even 5$d$ metals. However, it was at some point realized that in itinerant metals the tendency is opposite, DFT nearly always {\it over}estimates the magnetic moment~\cite{larson2004magnetism}, because magnetic and nearly-magnetic metals are strongly fluctuating systems~\cite{moriya2012spin} and DFT is a mean field theory. Therefore, in good metals DFT+$U$ routinely worsens the agreement with the experiment with respect to their magnetic properties. The same behavior was found for the SCAN functional~\cite{fu2019dft}. Thus, there is little physical justification for using magnetic versions of either DFT+$U$ or SCAN to describe RuO$_2$.

Therefore, it is not surprising that SCAN AFM calculations fail to describe the phonon data overemphasizing the electron-phonon coupling-induced Kohn anomaly near the Z point. That DFT without $U$ has similar issues is probably a coincidence. It remains unclear why non-magnetic SCAN appears superior in reproducing both phonon and EFG measurements; systematic studies of SCAN performance regarding these two calculations is lacking and the community intuition is not well established yet.\vspace{-1.95em}

\section*{Conclusion}\vspace{-1.25em}

In conclusion, we provide evidence that RuO$_2$ is non-magnetic based on nuclear resonant experiments ($^{99}$Ru M\"ossbauer spectroscopy and nuclear forward scattering techniques) coupled with inelastic X-ray and neutron scattering experiments and density functional theory (DFT) calculations. Nuclear resonant experiments showed that RuO$_2$ has zero or, at best, a negligible hyperfine field. Inelastic X-ray scattering provides dispersions for the acoustic phonons along the $\Gamma-$Z and $\Gamma-$R directions, while inelastic neutron scattering provides powder averaged acoustic and optical phonons. In a direct comparison of the experimental phonon data with DFT calculated phonons using SCAN and DFT+$U$ functionals, we conclude that the lattice dynamics in RuO$_2$ is best reproduced with a zero magnetic moment on the Ru-atoms using SCAN, albeit governed more strongly by electronic correlations than magnetization. Our experimental results reveal that bulk RuO$_2$ is non-magnetic in nature and hence is not a good altermagnet as previously thought. However, the precise nature of the electronic correlations remains unclear and RuO$_2$ still appears to be a powerful and challenging test case for assessing calculation accuracy. Although RuO$_2$ is intrinsically non-magnetic, the application of doping or strain could still induce a magnetic moment, rendering it a promising candidate for spintronic applications.\vspace{-1.75em}

\section*{Methods}\vspace{-1.25em}

\textit{Crystal Growth} Single crystals of RuO$_2$ were grown using the chemical vapor transport method. Powders of RuO$_2$ and Ru were mixed with a molar ratio of 1:9. The growth was carried out in a tube furnace with flowing oxygen as the transport agent. The hot end was kept at 1320$^{\circ}$C and the growth lasted 10 days. Millimeter-sized samples with a range of morphologies, including bar and cuboidal shapes, were obtained at the cold section of the tube.\newline
\hspace*{2ex}\textit{M\"ossbauer Spectroscopy and Nuclear Forward Scattering} The $^{99}$Ru M\"ossbauer data in this work was obtained from Ref.~\cite{stievano1999mossbauer} and the Nuclear forward scattering data was obtained from Ref.~\cite{bessas2014nfs}. We used the Nexus code~\cite{nexus2024} to perform the simultaneous fitting of the M\"ossbauer data and nuclear forward scattering data. \newline
\hspace*{2ex}\textit{Inelastic X-ray Scattering (IXS)} We performed the IXS measurements at beamline 30-ID(B) of the Advanced Photon Source (APS) at Argonne National Laboratory. Single crystal RuO$_2$ was aligned in the (0, K, L) plane which covers the $\Gamma-$Z and $\Gamma-$R directions as shown in Fig.~\ref{fig2}~(b). The experimental data were fitted using a damped harmonic oscillator (DHO) function convoluted with the instrument resolution. The IXS phonon data as shown in Fig.~\ref{fig2}~(e-j) are peak centers of the fits, and the error bars are the FWHM of the DHO profile. The error of the data point (peak center) is small and within the size of the circle.\newline
\hspace*{2ex}\textit{Inelastic Neutron Scattering (INS)} We performed INS measurements on the wide angle chopper spectrometer (ARCS) at the Spallation Neutron Source (SNS) of Oak Ridge National Laboratory. Powder RuO$_2$ samples (5 gram, 98\% Colonial Metals Inc.) were loaded in an aluminum sample can and the inelastic spectra were measured at room temperature with an incident energy of 150 meV. An empty aluminum sample can was also measured with the same incident energy, and this was used to make a background subtraction. For extracting the integrated $\chi''(E)$, the $Q$ integration range was chosen between 7 -- 12 \AA$^{-1}$ to avoid the low intensities on the low-$Q$ regions and to avoid the parabolic cut-off (high$-Q$ and high$-E$ region) due to the neutron kinetics. \newline
\hspace*{2ex}\textit{Density Functional Theory calculations} We performed DFT calculations using the projector augmented wave (PAW) method~\cite{Kresse_1999, Blöchl_1994} in VASP (v6.3.2)~\cite{Kresse_1996_1, Kresse_1996_2} with Ru (5s$^1$4d$^7$) and O (2s$^2$2p$^4$) potentials. All calculations used a plane-wave energy cuttoff of 600 eV and an $11\times11\times15$ Monkhorst-Pack k-point mesh. Structural relaxations were performed until the forces on atoms were below $10^{-4}$ eV/\AA\ and the energy threshold reached $10^{-6}$ eV. To assess the influence of exchange-correlation, we compared calculations employing (a) PBE~\cite{PBE_1996} with a Hubbard $U$ correction ($U_{\mathrm{eff}}$ = 0 and 2 eV) applied to the Ru $d$-orbitals, and (b) r$^2$SCAN meta-GGA functional with the rVV10 nonlocal correlation functional which has been shown to yield reliable predictions of both energetic and structural properties across a range of bonding environments~\cite{r2SCAN}. Phonon band structures were obtained from second-order interatomic force constants (IFCs), calculated using the finite displacement method in Phonopy~\cite{phonopy_2023}. For all phonon calculations, a 3$\times$3$\times$4 supercells (216 atoms) with a 2$\times$2$\times$2 Monkhorst-Pack $k$-point grid were employed.\newline 
\hspace*{2ex}The dynamic structure factor $S(Q,E)$ calculated via the OCLIMAX code is based on the one-phonon coherent inelastic scattering limit as described in Ref.~\cite{oclimax2019jctc}. The dynamic structure factor for inelastic X-ray scattering one-phonon coherent inelastic scattering takes the same form as in the case of neutrons, where the structure factor in IXS is given by the atomic form factor, $f_j(Q)$, instead of the neutron scattering length density, $b_j$. The relation between $\chi''(Q,E)$ and $S(Q,E)$ is given by the fluctuation dissipation theorem~\cite{RKubo_1966}. Note that the OCLIMAX calculations of $\chi''(E)$ from INS include contributions from multiphonon processes, i.e. the convolution of the one-phonon contribution with itself (and higher-order terms).\vspace{-1.75em}

\begin{acknowledgments}
\vspace{-1.25em}We thank L. Stievano and D. Bessas for provision of datafiles of initial M\"ossbauer spectroscopy and NFS of RuO$_2$, respectively, J. Hong for discussion on the phonon band structure in RuO$_{2}$ DFT calculations. D. Mandrus, R. Valenti, I. Sergueev, J. Yan, and P. Kent are acknowledged for helpful discussions. Research was supported by the U.S. Department of Energy, Office of Science, Basic Energy Sciences, Materials Sciences and Engineering Division (BES-MSED). L.B. acknowledges funding from the cluster of excellence “CUI: Advanced imaging of Matter”  of the Deutsche Forschungsgemeinschaft (DFG) – EXC 2056 – project ID 390715994 and support from DESY (Hamburg, Germany), a member of the Helmholtz Association HGF. I.I.M. was supported by Army Research Office under Cooperative Agreement Number W911NF-22-2-0173. H.Z. (crystal growth) acknowledges the support from the U.S. Department of Energy, BES-MSED with Grant No.DE-SC0020254. B.A.F. ($\mu$SR and neutron analysis) was supported by the U.S. Department of Energy, Office of Science, Basic Energy Science through Award No. DE-SC0021134. Calculations used resources of the Compute and Data Environment for Science (CADES) at the Oak Ridge National Laboratory, which is supported by the Office of Science of the U.S. Department of Energy under Contract No. DE-AC05-00OR22725 and the National Energy Research Scientific Computing Center (NERSC), a Department of Energy Office of Science User Facility using NERSC Award BES-ERCAP-m1057. A portion of this research used resources at the Advanced Photon Source, a DOE Office of Science User Facility operated by Argonne National Laboratory. A portion of this research used resources at the Spallation Neutron Source, a DOE Office of Science User Facility operated by the Oak Ridge National Laboratory. The beam time was allocated to ARCS on proposal number IPTS-17267.1.\vspace{-1.5em}
\end{acknowledgments}

\subsection*{Author Contributions}\vspace{-1.25em}
GY performed the analysis of M\"ossbauer, NFS, IXS, INS data with the help of RPH, LB, and YC. PRR, IIM, LRL, DSP, VRC performed and analyzed DFT calculations and provide theory input. HZ grew single crystalline RuO$_2$. JDB, DB, AS performed the IXS experiment. JDB, MEM, and DA performed the INS experiments. BAF contributed $\mu$SR and neutron analysis. All authors contributed to writing and editing of the manuscript.\vspace{-1.5em}

\subsection*{Competing Interests}\vspace{-1.25em}
All authors declare no financial or non-financial competing interests.\vspace{-1.5em}

\subsection*{Data Availibility}\vspace{-1.25em}
The data presented in this study are available at \href{https://doi.org/10.14461/oncat.data/2563163}{https://doi.org/10.14461/oncat.data/2563163}\vspace{-1.5em}

\subsection*{Additional information}\vspace{-1.25em}\phantomsection
\textbf{Supplemental information}\label{suppinfo} The online version contains supplementary material available at https://doi.org/xxx \vspace{-1.5em}

\bibliographystyle{apsrev4-2}
\bibliography{main.bib}

\begin{thebibliography}{55}%
\makeatletter
\providecommand \@ifxundefined [1]{%
 \@ifx{#1\undefined}
}%
\providecommand \@ifnum [1]{%
 \ifnum #1\expandafter \@firstoftwo
 \else \expandafter \@secondoftwo
 \fi
}%
\providecommand \@ifx [1]{%
 \ifx #1\expandafter \@firstoftwo
 \else \expandafter \@secondoftwo
 \fi
}%
\providecommand \natexlab [1]{#1}%
\providecommand \enquote  [1]{``#1''}%
\providecommand \bibnamefont  [1]{#1}%
\providecommand \bibfnamefont [1]{#1}%
\providecommand \citenamefont [1]{#1}%
\providecommand \href@noop [0]{\@secondoftwo}%
\providecommand \href [0]{\begingroup \@sanitize@url \@href}%
\providecommand \@href[1]{\@@startlink{#1}\@@href}%
\providecommand \@@href[1]{\endgroup#1\@@endlink}%
\providecommand \@sanitize@url [0]{\catcode `\\12\catcode `\$12\catcode
  `\&12\catcode `\#12\catcode `\^12\catcode `\_12\catcode `\%12\relax}%
\providecommand \@@startlink[1]{}%
\providecommand \@@endlink[0]{}%
\providecommand \url  [0]{\begingroup\@sanitize@url \@url }%
\providecommand \@url [1]{\endgroup\@href {#1}{\urlprefix }}%
\providecommand \urlprefix  [0]{URL }%
\providecommand \Eprint [0]{\href }%
\providecommand \doibase [0]{https://doi.org/}%
\providecommand \selectlanguage [0]{\@gobble}%
\providecommand \bibinfo  [0]{\@secondoftwo}%
\providecommand \bibfield  [0]{\@secondoftwo}%
\providecommand \translation [1]{[#1]}%
\providecommand \BibitemOpen [0]{}%
\providecommand \bibitemStop [0]{}%
\providecommand \bibitemNoStop [0]{.\EOS\space}%
\providecommand \EOS [0]{\spacefactor3000\relax}%
\providecommand \BibitemShut  [1]{\csname bibitem#1\endcsname}%
\let\auto@bib@innerbib\@empty
\bibitem [{\citenamefont {\ifmmode~\check{S}\else \v{S}\fi{}mejkal}\ \emph
  {et~al.}(2022{\natexlab{a}})\citenamefont {\ifmmode~\check{S}\else
  \v{S}\fi{}mejkal}, \citenamefont {Sinova},\ and\ \citenamefont
  {Jungwirth}}]{libor2022altermagnet}%
  \BibitemOpen
  \bibfield  {author} {\bibinfo {author} {\bibfnamefont {L.}~\bibnamefont
  {\ifmmode~\check{S}\else \v{S}\fi{}mejkal}}, \bibinfo {author} {\bibfnamefont
  {J.}~\bibnamefont {Sinova}},\ and\ \bibinfo {author} {\bibfnamefont
  {T.}~\bibnamefont {Jungwirth}},\ }\href
  {https://doi.org/10.1103/PhysRevX.12.040501} {\bibfield  {journal} {\bibinfo
  {journal} {Phys. Rev. X}\ }\textbf {\bibinfo {volume} {12}},\ \bibinfo
  {pages} {040501} (\bibinfo {year} {2022}{\natexlab{a}})}\BibitemShut
  {NoStop}%
\bibitem [{\citenamefont {Chen}\ \emph {et~al.}(2014)\citenamefont {Chen},
  \citenamefont {Niu},\ and\ \citenamefont {MacDonald}}]{chen2014AHEinNCM}%
  \BibitemOpen
  \bibfield  {author} {\bibinfo {author} {\bibfnamefont {H.}~\bibnamefont
  {Chen}}, \bibinfo {author} {\bibfnamefont {Q.}~\bibnamefont {Niu}},\ and\
  \bibinfo {author} {\bibfnamefont {A.~H.}\ \bibnamefont {MacDonald}},\ }\href
  {https://doi.org/10.1103/PhysRevLett.112.017205} {\bibfield  {journal}
  {\bibinfo  {journal} {Phys. Rev. Lett.}\ }\textbf {\bibinfo {volume} {112}},\
  \bibinfo {pages} {017205} (\bibinfo {year} {2014})}\BibitemShut {NoStop}%
\bibitem [{\citenamefont {Mazin\text{,~et. al.}}(2021)}]{igor2021prediction}%
  \BibitemOpen
  \bibfield  {author} {\bibinfo {author} {\bibfnamefont {I.~I.}\ \bibnamefont
  {Mazin\text{,~et. al.}}},\ }\href {https://doi.org/10.1073/pnas.2108924118}
  {\bibfield  {journal} {\bibinfo  {journal} {Proc. Natl. Acad. Sci. U.S.A.}\
  }\textbf {\bibinfo {volume} {118}},\ \bibinfo {pages} {e2108924118} (\bibinfo
  {year} {2021})}\BibitemShut {NoStop}%
\bibitem [{\citenamefont {Gonzalez~Betancourt\textit{,~et.
  al.}}(2023)}]{betancourt2023MnTe}%
  \BibitemOpen
  \bibfield  {author} {\bibinfo {author} {\bibfnamefont {R.~D.}\ \bibnamefont
  {Gonzalez~Betancourt\textit{,~et. al.}}},\ }\href
  {https://doi.org/10.1103/PhysRevLett.130.036702} {\bibfield  {journal}
  {\bibinfo  {journal} {Phys. Rev. Lett.}\ }\textbf {\bibinfo {volume} {130}},\
  \bibinfo {pages} {036702} (\bibinfo {year} {2023})}\BibitemShut {NoStop}%
\bibitem [{\citenamefont {Krempask{\'y}\textit{,~et.
  al.}}(2024)}]{krempasky2024altermagnetic}%
  \BibitemOpen
  \bibfield  {author} {\bibinfo {author} {\bibfnamefont {J.}~\bibnamefont
  {Krempask{\'y}\textit{,~et. al.}}},\ }\href
  {https://doi.org/10.1038/s41586-023-06907-7} {\bibfield  {journal} {\bibinfo
  {journal} {Nature}\ }\textbf {\bibinfo {volume} {626}},\ \bibinfo {pages}
  {517} (\bibinfo {year} {2024})}\BibitemShut {NoStop}%
\bibitem [{\citenamefont {Osumi\textit{,~et.
  al.}}(2024)}]{osumi2024observation}%
  \BibitemOpen
  \bibfield  {author} {\bibinfo {author} {\bibfnamefont {T.}~\bibnamefont
  {Osumi\textit{,~et. al.}}},\ }\href
  {https://doi.org/10.1103/PhysRevB.109.115102} {\bibfield  {journal} {\bibinfo
   {journal} {Phys. Rev. B}\ }\textbf {\bibinfo {volume} {109}},\ \bibinfo
  {pages} {115102} (\bibinfo {year} {2024})}\BibitemShut {NoStop}%
\bibitem [{\citenamefont {\ifmmode~\check{S}\else \v{S}\fi{}mejkal}\ \emph
  {et~al.}(2022{\natexlab{b}})\citenamefont {\ifmmode~\check{S}\else
  \v{S}\fi{}mejkal}, \citenamefont {Sinova},\ and\ \citenamefont
  {Jungwirth}}]{libor2022symmetry}%
  \BibitemOpen
  \bibfield  {author} {\bibinfo {author} {\bibfnamefont {L.}~\bibnamefont
  {\ifmmode~\check{S}\else \v{S}\fi{}mejkal}}, \bibinfo {author} {\bibfnamefont
  {J.}~\bibnamefont {Sinova}},\ and\ \bibinfo {author} {\bibfnamefont
  {T.}~\bibnamefont {Jungwirth}},\ }\href
  {https://doi.org/10.1103/PhysRevX.12.031042} {\bibfield  {journal} {\bibinfo
  {journal} {Phys. Rev. X}\ }\textbf {\bibinfo {volume} {12}},\ \bibinfo
  {pages} {031042} (\bibinfo {year} {2022}{\natexlab{b}})}\BibitemShut
  {NoStop}%
\bibitem [{\citenamefont {Olejník\textit{,~et. al.}}(2018)}]{kamil2018thz}%
  \BibitemOpen
  \bibfield  {author} {\bibinfo {author} {\bibfnamefont {K.}~\bibnamefont
  {Olejník\textit{,~et. al.}}},\ }\href
  {https://doi.org/10.1126/sciadv.aar3566} {\bibfield  {journal} {\bibinfo
  {journal} {Sci. Adv.}\ }\textbf {\bibinfo {volume} {4}},\ \bibinfo {pages}
  {eaar3566} (\bibinfo {year} {2018})}\BibitemShut {NoStop}%
\bibitem [{\citenamefont {Jungwirth}\ \emph {et~al.}(2016)\citenamefont
  {Jungwirth}, \citenamefont {Marti}, \citenamefont {Wadley},\ and\
  \citenamefont {Wunderlich}}]{tomas2016afms}%
  \BibitemOpen
  \bibfield  {author} {\bibinfo {author} {\bibfnamefont {T.}~\bibnamefont
  {Jungwirth}}, \bibinfo {author} {\bibfnamefont {X.}~\bibnamefont {Marti}},
  \bibinfo {author} {\bibfnamefont {P.}~\bibnamefont {Wadley}},\ and\ \bibinfo
  {author} {\bibfnamefont {J.}~\bibnamefont {Wunderlich}},\ }\href
  {https://doi.org/10.1038/nnano.2016.18} {\bibfield  {journal} {\bibinfo
  {journal} {Nat. Nanotechnol.}\ }\textbf {\bibinfo {volume} {11}},\ \bibinfo
  {pages} {231} (\bibinfo {year} {2016})}\BibitemShut {NoStop}%
\bibitem [{\citenamefont {Baltz\textit{,~et. al.}}(2018)}]{baltz2018afm}%
  \BibitemOpen
  \bibfield  {author} {\bibinfo {author} {\bibfnamefont {V.}~\bibnamefont
  {Baltz\textit{,~et. al.}}},\ }\href
  {https://doi.org/10.1103/RevModPhys.90.015005} {\bibfield  {journal}
  {\bibinfo  {journal} {Rev. Mod. Phys.}\ }\textbf {\bibinfo {volume} {90}},\
  \bibinfo {pages} {015005} (\bibinfo {year} {2018})}\BibitemShut {NoStop}%
\bibitem [{\citenamefont {Dal~Din}\ \emph {et~al.}(2024)\citenamefont
  {Dal~Din}, \citenamefont {Amin}, \citenamefont {Wadley},\ and\ \citenamefont
  {Edmonds}}]{Dal2024AFM}%
  \BibitemOpen
  \bibfield  {author} {\bibinfo {author} {\bibfnamefont {A.}~\bibnamefont
  {Dal~Din}}, \bibinfo {author} {\bibfnamefont {O.~J.}\ \bibnamefont {Amin}},
  \bibinfo {author} {\bibfnamefont {P.}~\bibnamefont {Wadley}},\ and\ \bibinfo
  {author} {\bibfnamefont {K.~W.}\ \bibnamefont {Edmonds}},\ }\href
  {https://doi.org/10.1038/s44306-024-00029-0} {\bibfield  {journal} {\bibinfo
  {journal} {npj Spintronics}\ }\textbf {\bibinfo {volume} {2}},\ \bibinfo
  {pages} {25} (\bibinfo {year} {2024})}\BibitemShut {NoStop}%
\bibitem [{\citenamefont {Šmejkal}\ \emph {et~al.}(2020)\citenamefont
  {Šmejkal}, \citenamefont {González-Hernández}, \citenamefont {Jungwirth},\
  and\ \citenamefont {Sinova}}]{libor2020che}%
  \BibitemOpen
  \bibfield  {author} {\bibinfo {author} {\bibfnamefont {L.}~\bibnamefont
  {Šmejkal}}, \bibinfo {author} {\bibfnamefont {R.}~\bibnamefont
  {González-Hernández}}, \bibinfo {author} {\bibfnamefont {T.}~\bibnamefont
  {Jungwirth}},\ and\ \bibinfo {author} {\bibfnamefont {J.}~\bibnamefont
  {Sinova}},\ }\href {https://doi.org/10.1126/sciadv.aaz8809} {\bibfield
  {journal} {\bibinfo  {journal} {Sci. Adv.}\ }\textbf {\bibinfo {volume}
  {6}},\ \bibinfo {pages} {eaaz8809} (\bibinfo {year} {2020})}\BibitemShut
  {NoStop}%
\bibitem [{\citenamefont {Feng\textit{,~et. al.}}(2022)}]{Feng2022anomalous}%
  \BibitemOpen
  \bibfield  {author} {\bibinfo {author} {\bibfnamefont {Z.}~\bibnamefont
  {Feng\textit{,~et. al.}}},\ }\href
  {https://doi.org/10.1038/s41928-022-00866-z} {\bibfield  {journal} {\bibinfo
  {journal} {Nat. Electron.}\ }\textbf {\bibinfo {volume} {5}},\ \bibinfo
  {pages} {735} (\bibinfo {year} {2022})}\BibitemShut {NoStop}%
\bibitem [{\citenamefont {{\v{S}}mejkal\textit{,~et.
  al.}}(2023)}]{libor2023chiral}%
  \BibitemOpen
  \bibfield  {author} {\bibinfo {author} {\bibfnamefont {L.}~\bibnamefont
  {{\v{S}}mejkal\textit{,~et. al.}}},\ }\href
  {https://doi.org/10.1103/PhysRevLett.131.256703} {\bibfield  {journal}
  {\bibinfo  {journal} {Phys. Rev. Lett.}\ }\textbf {\bibinfo {volume} {131}},\
  \bibinfo {pages} {256703} (\bibinfo {year} {2023})}\BibitemShut {NoStop}%
\bibitem [{\citenamefont {Zhou\textit{,~et. al.}}(2024)}]{zhou2024crystal}%
  \BibitemOpen
  \bibfield  {author} {\bibinfo {author} {\bibfnamefont {X.}~\bibnamefont
  {Zhou\textit{,~et. al.}}},\ }\href
  {https://doi.org/10.1103/PhysRevLett.132.056701} {\bibfield  {journal}
  {\bibinfo  {journal} {Phys. Rev. Lett.}\ }\textbf {\bibinfo {volume} {132}},\
  \bibinfo {pages} {056701} (\bibinfo {year} {2024})}\BibitemShut {NoStop}%
\bibitem [{\citenamefont {Fedchenko\textit{,~et. al.}}(2024)}]{olena2024obsTR}%
  \BibitemOpen
  \bibfield  {author} {\bibinfo {author} {\bibfnamefont {O.}~\bibnamefont
  {Fedchenko\textit{,~et. al.}}},\ }\href
  {https://doi.org/10.1126/sciadv.adj4883} {\bibfield  {journal} {\bibinfo
  {journal} {Sci. Adv.}\ }\textbf {\bibinfo {volume} {10}},\ \bibinfo {pages}
  {eadj4883} (\bibinfo {year} {2024})}\BibitemShut {NoStop}%
\bibitem [{\citenamefont {Guo\textit{,~et. al.}}(2024)}]{yaqin2024direct}%
  \BibitemOpen
  \bibfield  {author} {\bibinfo {author} {\bibfnamefont {Y.}~\bibnamefont
  {Guo\textit{,~et. al.}}},\ }\href
  {https://doi.org/https://doi.org/10.1002/advs.202400967} {\bibfield
  {journal} {\bibinfo  {journal} {Adv. Sci.}\ }\textbf {\bibinfo {volume}
  {11}},\ \bibinfo {pages} {2400967} (\bibinfo {year} {2024})}\BibitemShut
  {NoStop}%
\bibitem [{\citenamefont {Smolyanyuk}\ \emph {et~al.}(2024)\citenamefont
  {Smolyanyuk}, \citenamefont {Mazin}, \citenamefont {Garcia-Gassull},\ and\
  \citenamefont {Valent\'{\i}}}]{smolyanyuk2024fragility}%
  \BibitemOpen
  \bibfield  {author} {\bibinfo {author} {\bibfnamefont {A.}~\bibnamefont
  {Smolyanyuk}}, \bibinfo {author} {\bibfnamefont {I.~I.}\ \bibnamefont
  {Mazin}}, \bibinfo {author} {\bibfnamefont {L.}~\bibnamefont
  {Garcia-Gassull}},\ and\ \bibinfo {author} {\bibfnamefont {R.}~\bibnamefont
  {Valent\'{\i}}},\ }\href {https://doi.org/10.1103/PhysRevB.109.134424}
  {\bibfield  {journal} {\bibinfo  {journal} {Phys. Rev. B}\ }\textbf {\bibinfo
  {volume} {109}},\ \bibinfo {pages} {134424} (\bibinfo {year}
  {2024})}\BibitemShut {NoStop}%
\bibitem [{\citenamefont {Berlijn\textit{,~et.
  al.}}(2017)}]{berlijn2017neutron}%
  \BibitemOpen
  \bibfield  {author} {\bibinfo {author} {\bibfnamefont {T.}~\bibnamefont
  {Berlijn\textit{,~et. al.}}},\ }\href
  {https://doi.org/10.1103/PhysRevLett.118.077201} {\bibfield  {journal}
  {\bibinfo  {journal} {Phys. Rev. Lett.}\ }\textbf {\bibinfo {volume} {118}},\
  \bibinfo {pages} {077201} (\bibinfo {year} {2017})}\BibitemShut {NoStop}%
\bibitem [{\citenamefont {Ke{\ss}ler\textit{,~et.
  al.}}(2024)}]{kessler2024absence}%
  \BibitemOpen
  \bibfield  {author} {\bibinfo {author} {\bibfnamefont {P.}~\bibnamefont
  {Ke{\ss}ler\textit{,~et. al.}}},\ }\href
  {https://doi.org/10.1038/s44306-024-00055-y} {\bibfield  {journal} {\bibinfo
  {journal} {npj Spintronics}\ }\textbf {\bibinfo {volume} {2}},\ \bibinfo
  {pages} {50} (\bibinfo {year} {2024})}\BibitemShut {NoStop}%
\bibitem [{\citenamefont {Hiraishi\textit{,~et.
  al.}}(2024)}]{Hiraishi2024nonmagnetic}%
  \BibitemOpen
  \bibfield  {author} {\bibinfo {author} {\bibfnamefont {M.}~\bibnamefont
  {Hiraishi\textit{,~et. al.}}},\ }\href
  {https://doi.org/10.1103/PhysRevLett.132.166702} {\bibfield  {journal}
  {\bibinfo  {journal} {Phys. Rev. Lett.}\ }\textbf {\bibinfo {volume} {132}},\
  \bibinfo {pages} {166702} (\bibinfo {year} {2024})}\BibitemShut {NoStop}%
\bibitem [{\citenamefont {Liu\textit{,~et.
  al.}}(2024)}]{jianyu2024absencespinsplit}%
  \BibitemOpen
  \bibfield  {author} {\bibinfo {author} {\bibfnamefont {J.}~\bibnamefont
  {Liu\textit{,~et. al.}}},\ }\href
  {https://doi.org/10.1103/PhysRevLett.133.176401} {\bibfield  {journal}
  {\bibinfo  {journal} {Phys. Rev. Lett.}\ }\textbf {\bibinfo {volume} {133}},\
  \bibinfo {pages} {176401} (\bibinfo {year} {2024})}\BibitemShut {NoStop}%
\bibitem [{\citenamefont {Lovesey}\ \emph {et~al.}(2023)\citenamefont
  {Lovesey}, \citenamefont {Khalyavin},\ and\ \citenamefont {van~der
  Laan}}]{lovesey2023magstr}%
  \BibitemOpen
  \bibfield  {author} {\bibinfo {author} {\bibfnamefont {S.~W.}\ \bibnamefont
  {Lovesey}}, \bibinfo {author} {\bibfnamefont {D.~D.}\ \bibnamefont
  {Khalyavin}},\ and\ \bibinfo {author} {\bibfnamefont {G.}~\bibnamefont
  {van~der Laan}},\ }\href {https://doi.org/10.1103/PhysRevB.108.L121103}
  {\bibfield  {journal} {\bibinfo  {journal} {Phys. Rev. B}\ }\textbf {\bibinfo
  {volume} {108}},\ \bibinfo {pages} {L121103} (\bibinfo {year}
  {2023})}\BibitemShut {NoStop}%
\bibitem [{\citenamefont {Bolzan}\ \emph {et~al.}(1997)\citenamefont {Bolzan},
  \citenamefont {Fong}, \citenamefont {Kennedy},\ and\ \citenamefont
  {Howard}}]{bolzan1997structural}%
  \BibitemOpen
  \bibfield  {author} {\bibinfo {author} {\bibfnamefont {A.~A.}\ \bibnamefont
  {Bolzan}}, \bibinfo {author} {\bibfnamefont {C.}~\bibnamefont {Fong}},
  \bibinfo {author} {\bibfnamefont {B.~J.}\ \bibnamefont {Kennedy}},\ and\
  \bibinfo {author} {\bibfnamefont {C.~J.}\ \bibnamefont {Howard}},\ }\href
  {https://doi.org/https://doi.org/10.1107/S0108768197001468} {\bibfield
  {journal} {\bibinfo  {journal} {Acta Crystallogr. Sect. B}\ }\textbf
  {\bibinfo {volume} {53}},\ \bibinfo {pages} {373} (\bibinfo {year}
  {1997})}\BibitemShut {NoStop}%
\bibitem [{\citenamefont {Jacob}\ \emph {et~al.}(2000)\citenamefont {Jacob},
  \citenamefont {Mishra},\ and\ \citenamefont {Waseda}}]{jacob2000refinement}%
  \BibitemOpen
  \bibfield  {author} {\bibinfo {author} {\bibfnamefont {K.~T.}\ \bibnamefont
  {Jacob}}, \bibinfo {author} {\bibfnamefont {S.}~\bibnamefont {Mishra}},\ and\
  \bibinfo {author} {\bibfnamefont {Y.}~\bibnamefont {Waseda}},\ }\href
  {https://doi.org/https://doi.org/10.1111/j.1151-2916.2000.tb01459.x}
  {\bibfield  {journal} {\bibinfo  {journal} {J. Am. Ceram. Soc.}\ }\textbf
  {\bibinfo {volume} {83}},\ \bibinfo {pages} {1745} (\bibinfo {year}
  {2000})}\BibitemShut {NoStop}%
\bibitem [{\citenamefont {Stievano}\ \emph {et~al.}(1999)\citenamefont
  {Stievano}, \citenamefont {Calogero}, \citenamefont {Wagner}, \citenamefont
  {Galvagno},\ and\ \citenamefont {Milone}}]{stievano1999mossbauer}%
  \BibitemOpen
  \bibfield  {author} {\bibinfo {author} {\bibfnamefont {L.}~\bibnamefont
  {Stievano}}, \bibinfo {author} {\bibfnamefont {S.}~\bibnamefont {Calogero}},
  \bibinfo {author} {\bibfnamefont {F.~E.}\ \bibnamefont {Wagner}}, \bibinfo
  {author} {\bibfnamefont {S.}~\bibnamefont {Galvagno}},\ and\ \bibinfo
  {author} {\bibfnamefont {C.}~\bibnamefont {Milone}},\ }\href
  {https://doi.org/10.1021/jp990420e} {\bibfield  {journal} {\bibinfo
  {journal} {J. Phys. Chem. B}\ }\textbf {\bibinfo {volume} {103}},\ \bibinfo
  {pages} {9545} (\bibinfo {year} {1999})}\BibitemShut {NoStop}%
\bibitem [{\citenamefont {Bessas\textit{,~et. al.}}(2014)}]{bessas2014nfs}%
  \BibitemOpen
  \bibfield  {author} {\bibinfo {author} {\bibfnamefont {D.}~\bibnamefont
  {Bessas\textit{,~et. al.}}},\ }\href
  {https://doi.org/10.1103/PhysRevLett.113.147601} {\bibfield  {journal}
  {\bibinfo  {journal} {Phys. Rev. Lett.}\ }\textbf {\bibinfo {volume} {113}},\
  \bibinfo {pages} {147601} (\bibinfo {year} {2014})}\BibitemShut {NoStop}%
\bibitem [{\citenamefont {Kiefer\textit{,~et. al.}}(2025)}]{Kiefer2025}%
  \BibitemOpen
  \bibfield  {author} {\bibinfo {author} {\bibfnamefont {L.}~\bibnamefont
  {Kiefer\textit{,~et. al.}}},\ }\href
  {https://doi.org/10.1088/1361-648X/adad2a} {\bibfield  {journal} {\bibinfo
  {journal} {J. Phys. Condens. Matter}\ }\textbf {\bibinfo {volume} {37}},\
  \bibinfo {pages} {135801} (\bibinfo {year} {2025})}\BibitemShut {NoStop}%
\bibitem [{\citenamefont {Ko\textit{,~et. al.}}(2018)}]{ko2018understanding}%
  \BibitemOpen
  \bibfield  {author} {\bibinfo {author} {\bibfnamefont {D.-S.}\ \bibnamefont
  {Ko\textit{,~et. al.}}},\ }\href {https://doi.org/10.1038/s41427-018-0020-y}
  {\bibfield  {journal} {\bibinfo  {journal} {NPG Asia Mater.}\ }\textbf
  {\bibinfo {volume} {10}},\ \bibinfo {pages} {266} (\bibinfo {year}
  {2018})}\BibitemShut {NoStop}%
\bibitem [{\citenamefont {Gregory\textit{,~et.
  al.}}(2022)}]{gregory2022strain}%
  \BibitemOpen
  \bibfield  {author} {\bibinfo {author} {\bibfnamefont {B.~Z.}\ \bibnamefont
  {Gregory\textit{,~et. al.}}},\ }\href
  {https://doi.org/10.1103/PhysRevB.106.195135} {\bibfield  {journal} {\bibinfo
   {journal} {Phys. Rev. B}\ }\textbf {\bibinfo {volume} {106}},\ \bibinfo
  {pages} {195135} (\bibinfo {year} {2022})}\BibitemShut {NoStop}%
\bibitem [{\citenamefont {Chaudhary}\ \emph {et~al.}(2024)\citenamefont
  {Chaudhary}, \citenamefont {Zagalskaya}, \citenamefont {Over},\ and\
  \citenamefont {Alexandrov}}]{payal2024strain}%
  \BibitemOpen
  \bibfield  {author} {\bibinfo {author} {\bibfnamefont {P.}~\bibnamefont
  {Chaudhary}}, \bibinfo {author} {\bibfnamefont {A.}~\bibnamefont
  {Zagalskaya}}, \bibinfo {author} {\bibfnamefont {H.}~\bibnamefont {Over}},\
  and\ \bibinfo {author} {\bibfnamefont {V.}~\bibnamefont {Alexandrov}},\
  }\href {https://doi.org/https://doi.org/10.1002/celc.202300659} {\bibfield
  {journal} {\bibinfo  {journal} {ChemElectroChem}\ }\textbf {\bibinfo {volume}
  {11}},\ \bibinfo {pages} {e202300659} (\bibinfo {year} {2024})}\BibitemShut
  {NoStop}%
\bibitem [{\citenamefont {Wang\text{,~et.
  al.}}(2025)}]{wang2025robustanisotropicspinhall}%
  \BibitemOpen
  \bibfield  {author} {\bibinfo {author} {\bibfnamefont {Y.-C.}\ \bibnamefont
  {Wang\text{,~et. al.}}},\ }\href {https://arxiv.org/abs/2503.07985} {\bibinfo
  {title} {Robust anisotropic spin hall effect in rutile {RuO$_2$}}} (\bibinfo
  {year} {2025}),\ \Eprint {https://arxiv.org/abs/2503.07985}
  {arXiv:2503.07985} \BibitemShut {NoStop}%
\bibitem [{\citenamefont {Kistner}(1966)}]{kistner1966}%
  \BibitemOpen
  \bibfield  {author} {\bibinfo {author} {\bibfnamefont {O.~C.}\ \bibnamefont
  {Kistner}},\ }\href {https://doi.org/10.1103/PhysRev.144.1022} {\bibfield
  {journal} {\bibinfo  {journal} {Phys. Rev.}\ }\textbf {\bibinfo {volume}
  {144}},\ \bibinfo {pages} {1022} (\bibinfo {year} {1966})}\BibitemShut
  {NoStop}%
\bibitem [{\citenamefont {Long}\ and\ \citenamefont
  {Grandjean}(2013)}]{long2013mossbauer}%
  \BibitemOpen
  \bibfield  {author} {\bibinfo {author} {\bibfnamefont {G.~J.}\ \bibnamefont
  {Long}}\ and\ \bibinfo {author} {\bibfnamefont {F.}~\bibnamefont
  {Grandjean}},\ }\href@noop {} {\emph {\bibinfo {title} {M{\"o}ssbauer
  spectroscopy applied to magnetism and materials science}}},\ Vol.~\bibinfo
  {volume} {1}\ (\bibinfo  {publisher} {Springer Science \& Business Media},\
  \bibinfo {year} {2013})\BibitemShut {NoStop}%
\bibitem [{\citenamefont {Mukuda\textit{,~et.
  al.}}(1999)}]{mukuda1999spinfluct}%
  \BibitemOpen
  \bibfield  {author} {\bibinfo {author} {\bibfnamefont {H.}~\bibnamefont
  {Mukuda\textit{,~et. al.}}},\ }\href
  {https://doi.org/10.1103/PhysRevB.60.12279} {\bibfield  {journal} {\bibinfo
  {journal} {Phys. Rev. B}\ }\textbf {\bibinfo {volume} {60}},\ \bibinfo
  {pages} {12279} (\bibinfo {year} {1999})}\BibitemShut {NoStop}%
\bibitem [{\citenamefont {Bocklage}(2024)}]{nexus2024}%
  \BibitemOpen
  \bibfield  {author} {\bibinfo {author} {\bibfnamefont {L.}~\bibnamefont
  {Bocklage}},\ }\href {https://doi.org/10.5281/zenodo.13832946} {\bibinfo
  {title} {Nexus - nuclear elastic x-ray scattering universal software
  (version: 1.2.0)}} (\bibinfo {year} {2024}),\ \bibinfo {note} {{Z}enodo.
  {DOI}: 10.5281/zenodo.13832946 (last accessed: Jan, 2025)}\BibitemShut
  {NoStop}%
\bibitem [{\citenamefont {DeMarco\textit{,~et.
  al.}}(2000)}]{DeMarco2000temperature}%
  \BibitemOpen
  \bibfield  {author} {\bibinfo {author} {\bibfnamefont {M.}~\bibnamefont
  {DeMarco\textit{,~et. al.}}},\ }\href
  {https://doi.org/10.1103/PhysRevB.62.14297} {\bibfield  {journal} {\bibinfo
  {journal} {Phys. Rev. B}\ }\textbf {\bibinfo {volume} {62}},\ \bibinfo
  {pages} {14297} (\bibinfo {year} {2000})}\BibitemShut {NoStop}%
\bibitem [{\citenamefont {Dattagupta}\ and\ \citenamefont
  {Blume}(1974)}]{Dattagupta1974stochastic}%
  \BibitemOpen
  \bibfield  {author} {\bibinfo {author} {\bibfnamefont {S.}~\bibnamefont
  {Dattagupta}}\ and\ \bibinfo {author} {\bibfnamefont {M.}~\bibnamefont
  {Blume}},\ }\href {https://doi.org/10.1103/PhysRevB.10.4540} {\bibfield
  {journal} {\bibinfo  {journal} {Phys. Rev. B}\ }\textbf {\bibinfo {volume}
  {10}},\ \bibinfo {pages} {4540} (\bibinfo {year} {1974})}\BibitemShut
  {NoStop}%
\bibitem [{\citenamefont {Hermann\textit{,~et. al.}}(2006)}]{hermann2006}%
  \BibitemOpen
  \bibfield  {author} {\bibinfo {author} {\bibfnamefont {R.~P.}\ \bibnamefont
  {Hermann\textit{,~et. al.}}},\ }\href
  {https://doi.org/10.1103/PhysRevLett.97.017401} {\bibfield  {journal}
  {\bibinfo  {journal} {Phys. Rev. Lett.}\ }\textbf {\bibinfo {volume} {97}},\
  \bibinfo {pages} {017401} (\bibinfo {year} {2006})}\BibitemShut {NoStop}%
\bibitem [{\citenamefont {Hahn\textit{,~et. al.}}(2009)}]{hahn2009influence}%
  \BibitemOpen
  \bibfield  {author} {\bibinfo {author} {\bibfnamefont {S.~E.}\ \bibnamefont
  {Hahn\textit{,~et. al.}}},\ }\href
  {https://doi.org/10.1103/PhysRevB.79.220511} {\bibfield  {journal} {\bibinfo
  {journal} {Phys. Rev. B}\ }\textbf {\bibinfo {volume} {79}},\ \bibinfo
  {pages} {220511} (\bibinfo {year} {2009})}\BibitemShut {NoStop}%
\bibitem [{\citenamefont {Yildirim}(2009)}]{YILDIRIM2009425}%
  \BibitemOpen
  \bibfield  {author} {\bibinfo {author} {\bibfnamefont {T.}~\bibnamefont
  {Yildirim}},\ }\href
  {https://doi.org/https://doi.org/10.1016/j.physc.2009.03.038} {\bibfield
  {journal} {\bibinfo  {journal} {Phys. C: Supercond.}\ }\textbf {\bibinfo
  {volume} {469}},\ \bibinfo {pages} {425} (\bibinfo {year}
  {2009})}\BibitemShut {NoStop}%
\bibitem [{\citenamefont {Ning\textit{,~et. al.}}(2022)}]{r2SCAN}%
  \BibitemOpen
  \bibfield  {author} {\bibinfo {author} {\bibfnamefont {J.}~\bibnamefont
  {Ning\textit{,~et. al.}}},\ }\href
  {https://doi.org/10.1103/PhysRevB.106.075422} {\bibfield  {journal} {\bibinfo
   {journal} {Phys. Rev. B}\ }\textbf {\bibinfo {volume} {106}},\ \bibinfo
  {pages} {075422} (\bibinfo {year} {2022})}\BibitemShut {NoStop}%
\bibitem [{\citenamefont {Raghuvanshi\textit{,~et.
  al.}}(2025)}]{parul2025oso2}%
  \BibitemOpen
  \bibfield  {author} {\bibinfo {author} {\bibfnamefont {P.~R.}\ \bibnamefont
  {Raghuvanshi\textit{,~et. al.}}},\ }\href
  {https://doi.org/10.1103/PhysRevMaterials.9.034407} {\bibfield  {journal}
  {\bibinfo  {journal} {Phys. Rev. Mater.}\ }\textbf {\bibinfo {volume} {9}},\
  \bibinfo {pages} {034407} (\bibinfo {year} {2025})}\BibitemShut {NoStop}%
\bibitem [{\citenamefont {Cheng}\ \emph {et~al.}(2019)\citenamefont {Cheng},
  \citenamefont {Daemen}, \citenamefont {Kolesnikov},\ and\ \citenamefont
  {Ramirez-Cuesta}}]{oclimax2019jctc}%
  \BibitemOpen
  \bibfield  {author} {\bibinfo {author} {\bibfnamefont {Y.~Q.}\ \bibnamefont
  {Cheng}}, \bibinfo {author} {\bibfnamefont {L.~L.}\ \bibnamefont {Daemen}},
  \bibinfo {author} {\bibfnamefont {A.~I.}\ \bibnamefont {Kolesnikov}},\ and\
  \bibinfo {author} {\bibfnamefont {A.~J.}\ \bibnamefont {Ramirez-Cuesta}},\
  }\href {https://doi.org/10.1021/acs.jctc.8b01250} {\bibfield  {journal}
  {\bibinfo  {journal} {Journal of Chemical Theory and Computation}\ }\textbf
  {\bibinfo {volume} {15}},\ \bibinfo {pages} {1974} (\bibinfo {year}
  {2019})}\BibitemShut {NoStop}%
\bibitem [{\citenamefont {Anisimov}\ \emph {et~al.}(1991)\citenamefont
  {Anisimov}, \citenamefont {Zaanen},\ and\ \citenamefont
  {Andersen}}]{anisimov1991band}%
  \BibitemOpen
  \bibfield  {author} {\bibinfo {author} {\bibfnamefont {V.~I.}\ \bibnamefont
  {Anisimov}}, \bibinfo {author} {\bibfnamefont {J.}~\bibnamefont {Zaanen}},\
  and\ \bibinfo {author} {\bibfnamefont {O.~K.}\ \bibnamefont {Andersen}},\
  }\href {https://doi.org/10.1103/PhysRevB.44.943} {\bibfield  {journal}
  {\bibinfo  {journal} {Phys. Rev. B}\ }\textbf {\bibinfo {volume} {44}},\
  \bibinfo {pages} {943} (\bibinfo {year} {1991})}\BibitemShut {NoStop}%
\bibitem [{\citenamefont {Larson}\ \emph {et~al.}(2004)\citenamefont {Larson},
  \citenamefont {Mazin},\ and\ \citenamefont {Singh}}]{larson2004magnetism}%
  \BibitemOpen
  \bibfield  {author} {\bibinfo {author} {\bibfnamefont {P.}~\bibnamefont
  {Larson}}, \bibinfo {author} {\bibfnamefont {I.~I.}\ \bibnamefont {Mazin}},\
  and\ \bibinfo {author} {\bibfnamefont {D.~J.}\ \bibnamefont {Singh}},\ }\href
  {https://doi.org/10.1103/PhysRevB.69.064429} {\bibfield  {journal} {\bibinfo
  {journal} {Phys. Rev. B}\ }\textbf {\bibinfo {volume} {69}},\ \bibinfo
  {pages} {064429} (\bibinfo {year} {2004})}\BibitemShut {NoStop}%
\bibitem [{\citenamefont {Moriya}(2012)}]{moriya2012spin}%
  \BibitemOpen
  \bibfield  {author} {\bibinfo {author} {\bibfnamefont {T.}~\bibnamefont
  {Moriya}},\ }\href@noop {} {\emph {\bibinfo {title} {Spin fluctuations in
  itinerant electron magnetism}}},\ Vol.~\bibinfo {volume} {56}\ (\bibinfo
  {publisher} {Springer Science \& Business Media},\ \bibinfo {year}
  {2012})\BibitemShut {NoStop}%
\bibitem [{\citenamefont {Fu}\ and\ \citenamefont {Singh}(2019)}]{fu2019dft}%
  \BibitemOpen
  \bibfield  {author} {\bibinfo {author} {\bibfnamefont {Y.}~\bibnamefont
  {Fu}}\ and\ \bibinfo {author} {\bibfnamefont {D.~J.}\ \bibnamefont {Singh}},\
  }\href {https://doi.org/10.1103/PhysRevB.100.045126} {\bibfield  {journal}
  {\bibinfo  {journal} {Phys. Rev. B}\ }\textbf {\bibinfo {volume} {100}},\
  \bibinfo {pages} {045126} (\bibinfo {year} {2019})}\BibitemShut {NoStop}%
\bibitem [{\citenamefont {Kresse}\ and\ \citenamefont
  {Joubert}(1999)}]{Kresse_1999}%
  \BibitemOpen
  \bibfield  {author} {\bibinfo {author} {\bibfnamefont {G.}~\bibnamefont
  {Kresse}}\ and\ \bibinfo {author} {\bibfnamefont {D.}~\bibnamefont
  {Joubert}},\ }\href {https://doi.org/10.1103/PhysRevB.59.1758} {\bibfield
  {journal} {\bibinfo  {journal} {Phys. Rev. B}\ }\textbf {\bibinfo {volume}
  {59}},\ \bibinfo {pages} {1758} (\bibinfo {year} {1999})}\BibitemShut
  {NoStop}%
\bibitem [{\citenamefont {Bl\"ochl}(1994)}]{Blöchl_1994}%
  \BibitemOpen
  \bibfield  {author} {\bibinfo {author} {\bibfnamefont {P.~E.}\ \bibnamefont
  {Bl\"ochl}},\ }\href {https://doi.org/10.1103/PhysRevB.50.17953} {\bibfield
  {journal} {\bibinfo  {journal} {Phys. Rev. B}\ }\textbf {\bibinfo {volume}
  {50}},\ \bibinfo {pages} {17953} (\bibinfo {year} {1994})}\BibitemShut
  {NoStop}%
\bibitem [{\citenamefont {Kresse}\ and\ \citenamefont
  {Furthmüller}(1996)}]{Kresse_1996_1}%
  \BibitemOpen
  \bibfield  {author} {\bibinfo {author} {\bibfnamefont {G.}~\bibnamefont
  {Kresse}}\ and\ \bibinfo {author} {\bibfnamefont {J.}~\bibnamefont
  {Furthmüller}},\ }\href
  {https://doi.org/https://doi.org/10.1016/0927-0256(96)00008-0} {\bibfield
  {journal} {\bibinfo  {journal} {Comput. Mater. Sci.}\ }\textbf {\bibinfo
  {volume} {6}},\ \bibinfo {pages} {15} (\bibinfo {year} {1996})}\BibitemShut
  {NoStop}%
\bibitem [{\citenamefont {Kresse}\ and\ \citenamefont
  {Furthm\"uller}(1996)}]{Kresse_1996_2}%
  \BibitemOpen
  \bibfield  {author} {\bibinfo {author} {\bibfnamefont {G.}~\bibnamefont
  {Kresse}}\ and\ \bibinfo {author} {\bibfnamefont {J.}~\bibnamefont
  {Furthm\"uller}},\ }\href {https://doi.org/10.1103/PhysRevB.54.11169}
  {\bibfield  {journal} {\bibinfo  {journal} {Phys. Rev. B}\ }\textbf {\bibinfo
  {volume} {54}},\ \bibinfo {pages} {11169} (\bibinfo {year}
  {1996})}\BibitemShut {NoStop}%
\bibitem [{\citenamefont {Perdew}\ \emph {et~al.}(1996)\citenamefont {Perdew},
  \citenamefont {Burke},\ and\ \citenamefont {Ernzerhof}}]{PBE_1996}%
  \BibitemOpen
  \bibfield  {author} {\bibinfo {author} {\bibfnamefont {J.~P.}\ \bibnamefont
  {Perdew}}, \bibinfo {author} {\bibfnamefont {K.}~\bibnamefont {Burke}},\ and\
  \bibinfo {author} {\bibfnamefont {M.}~\bibnamefont {Ernzerhof}},\ }\href
  {https://doi.org/10.1103/PhysRevLett.77.3865} {\bibfield  {journal} {\bibinfo
   {journal} {Phys. Rev. Lett.}\ }\textbf {\bibinfo {volume} {77}},\ \bibinfo
  {pages} {3865} (\bibinfo {year} {1996})}\BibitemShut {NoStop}%
\bibitem [{\citenamefont {Togo}(2023)}]{phonopy_2023}%
  \BibitemOpen
  \bibfield  {author} {\bibinfo {author} {\bibfnamefont {A.}~\bibnamefont
  {Togo}},\ }\href {https://doi.org/10.7566/JPSJ.92.012001} {\bibfield
  {journal} {\bibinfo  {journal} {J. Phys. Soc. Jpn.}\ }\textbf {\bibinfo
  {volume} {92}},\ \bibinfo {pages} {012001} (\bibinfo {year}
  {2023})}\BibitemShut {NoStop}%
\bibitem [{\citenamefont {Kubo}(1966)}]{RKubo_1966}%
  \BibitemOpen
  \bibfield  {author} {\bibinfo {author} {\bibfnamefont {R.}~\bibnamefont
  {Kubo}},\ }\href {https://doi.org/10.1088/0034-4885/29/1/306} {\bibfield
  {journal} {\bibinfo  {journal} {Rep. Prog. Phys.}\ }\textbf {\bibinfo
  {volume} {29}},\ \bibinfo {pages} {255} (\bibinfo {year} {1966})}\BibitemShut
  {NoStop}%
\end{thebibliography}%
\end{document}